%% file: manuscript.tex
\documentclass[tighten, times, twocolumn]{aastex62}  
\input{top-apj}
\usepackage[inline]{enumitem}
\usepackage{upgreek}

\newcommand{\cc}{\,\textrm{cm}^{-3}}
\shorttitle{Synthetic Polarization Maps}
\shortauthors{Bino {\em et al.}}

\begin{document}

\title{Synthetic Polarization Maps of an Outflow Zone from Magnetohydrodynamic Simulations}

\author{Gianfranco Bino}
\affiliation{Department of Applied Mathematics, University of Western Ontario, London, ON, N6A 5B7, Canada.}
\affiliation{Toronto-Dominion Bank, Toronto, ON, M5K 1B1, Canada.}

\author{Shantanu Basu}
\affiliation{Department of Physics \& Astronomy, University of Western Ontario, London, ON, N6A 3K7, Canada.}
\affiliation{Institute for Earth \& Space Exploration, University of Western Ontario, London, ON, N6A 5B7, Canada.}

\author{Masahiro N. Machida}
\affiliation{Department of Earth and Planetary Sciences, Faculty of Sciences, Kyushu University, Fukuyoka  819-0395, Japan.}
\affiliation{Department of Physics \& Astronomy, University of Western Ontario, London, ON, N6A 3K7, Canada.}

\author{Aris Tritsis}
\affiliation{Department of Physics \& Astronomy, University of Western Ontario, London, ON, N6A 3K7, Canada.}
\affiliation{CITA National Fellow.}

\author{Mahmoud Sharkawi}
\affiliation{Department of Applied Mathematics, University of Western Ontario, London, ON, N6A 5B7, Canada.}

\author{Kundan Kadam}
\affiliation{Department of Physics \& Astronomy, University of Western Ontario, London, ON, N6A 3K7, Canada.}

\author{Indrani Das}
\affiliation{Department of Physics \& Astronomy, University of Western Ontario, London, ON, N6A 3K7, Canada.}


\email{gbino@uwo.ca}
\email{basu@uwo.ca}

\begin{abstract}
The canonical theory of star formation in a magnetized environment predicts the formation of hourglass-shaped magnetic fields during the prestellar collapse phase. In protostellar cores, recent observations reveal complex and strongly distorted magnetic fields in the inner regions that are sculpted by rotation and outflows. 
We conduct resistive, nonideal magnetohydrodynamic (MHD) simulations of a protostellar core and employ the radiative transfer code POLARIS to produce synthetic polarization segment maps.
Comparison of our mock-polarization maps based on the toroidal-dominated magnetic field in the outflow zone with the observed polarization vectors of SiO lines in Orion Source I shows a reasonable agreement when the magnetic axis is tilted at an angle $\theta = 15^{\circ}$ with respect to the plane-of-sky and if the SiO lines have a net polarization parallel to the local magnetic field.
Although the observed polarization is from SiO lines and our synthetic maps are due to polarized dust emission, a comparison is useful and allows us to resolve the ambiguity of whether the line polarization is parallel or perpendicular to the local magnetic field direction.

\end{abstract}


\section{Introduction}
\label{introduction}
The magnetic field plays an important role in the star formation process. The configuration of  magnetic fields (or magnetic field vectors) in star-forming cores can be inferred from polarization observations.
The observed polarization vectors (or segments) are expected to be either parallel or perpendicular to the magnetic field vectors, depending on the 
underlying physical process: thermal dust emission, dust extinction, or line emission. 
Recent polarization observations indicate that magnetic fields in prestellar and protostellar cores are well aligned across many scales \citep{girart06,girart09, PC16, ching16, alves18,lee18, maury18}. This indicates that the star-formation process is, at least in part, regulated
by the magnetic field rather than highly super-Alfv\'enic motions. If this is indeed the case, then the magnetic field morphology during the prestellar phase should resemble an hourglass configuration, as has been revealed in recent
observations with the Stratospheric Observatory for Infrared Astronomy (SOFIA) \citep{Chuss2019, Redaelli2019}.

However, many interesting polarization patterns have been observed that cannot be fitted neatly into the category of an hourglass pattern viewed from an extension of the equatorial plane. A majority of such objects also show evidence for outflows launched into the surrounding cloud at velocities $\sim$ a few tens of km s$^{-1}$ from an extended region $ \lesssim 100-1000$~au. The polarization patterns observed from these outflow zones
can be quite complex \citep[e.g.,][]{lee18,maury18,sadavoy2018,kwon2019,hir2020,hull2020}. Deciphering such patterns is crucial for understanding disk formation and outflow driving.

One of the most interesting targets is Orion Source I (hereafter Source I), a protostar located in the nearest high-mass star-forming region, Orion KL, in the Orion Nebula, at a distance $\simeq 420$ pc \citep{Zucker2020}.
For this source, a nearly edge-on disk with a size of $\sim50$\,au has been observed \citep{Reid2007,Goddi2011, Hirota2016,Plambeck2016, Ginsburg2018}.
Thus, Source I can serve as a testbed for studying the magnetic field both in the disk and the outflow zone in high-mass star formation regions.
Recently, \citet{hirota17} confirmed the rotation of the molecular outflow, in addition to the rotation of the disk. 
Then, using Bernoulli's theorem and the angular momentum conservation law, they estimated the outflow launching radius to be in the range $\sim5-25$\,au, indicating that the molecular outflow is directly driven from the disk surface, and not entrained by an unseen high-velocity jet. 
The rotation of molecular outflows has also been observed in other recent Atacama Large Millimeter/submillimeter Array (ALMA) observations \citep{bjerkeli16,tabone17,zhang18,mat21}.

The magnetic field in jet and outflow zones can also be inferred indirectly through molecular line polarization.
Recently, \citet{lee18} used ALMA observations to detect SiO line polarization in the $J = 8-7$ transition in the inner part of the HH 211 protostellar jet of a low-mass protostar. The polarization orientations were almost aligned with the jet axis and thus with the velocity flow axis.
Subsequently, \citet{hir2020} observed the polarized emission from SiO $J=1-0$ and $J=2-1$ lines within the outflow zone of Orion Source I, using the Very Large Array (VLA) and ALMA.
In these observations, the polarization vectors are well aligned, although they are neither parallel nor perpendicular to the outflow axis, i.e., the propagation direction of the outflow.
Since the polarization vectors are considered to be produced by the projected and integrated magnetic fields, it is non-trivial to infer the actual three-dimensional configuration of the magnetic field and the observed polarization pattern may depend on the inclination of the magnetic axis to the plane of sky. 
Thus, synthetic observations or synthetic polarization maps are necessary to understand the configuration of the magnetic field. 

Comparison of synthetic observations with the observed polarization segments is an indispensable tool for untangling the complex observations and relating them to underlying physical processes.
Analytic models for the magnetic field of prestellar or protostellar cores have been used to make synthetic maps that were compared with polarization vectors from individual objects \citep{Goncalves2008,Frau2011,Myers2020,binobasu2021}, but none of these models included rotation. Fitting the inner regions of protostellar cores, which contain highly twisted fields that also help to drive the outflow, requires comparison with a detailed numerical simulation of rotating magnetized collapse.
The first use of simulations to make synthetic polarization maps was carried out by \citet{tomisaka11}, who
utilized two-dimensional ideal MHD core collapse simulations.
In this study, the polarization degree and pattern both before and after the first core formation were discussed. 
\citet{reissl17} constructed synthetic observations using the data of a core collapse simulation. 
Their study focused on the dust alignment mechanism, however the simulations could not correctly resolve the disk and outflow-driving region because of the considerably large-sized sink (over 12\,au) at the center of the computational domain. 
In addition, the magnetic field strength adopted in their simulation was weak, with $\mu=26$, where $\mu$ is the mass-to-flux ratio normalized to the critical value for collapse. 
With such a weak magnetic field, a different mode of outflow appears in an ideal magnetohydrodynamic (MHD) simulation, as shown in \citet{tomisaka02}.
Despite being very useful, such past synthetic observations cannot be compared with the observations in the ALMA era, as the spatial resolution of the synthetic polarization maps in \citet{tomisaka11} and \citet{reissl17} is not sufficient.


Similarly, dust-emission polarimetry of the Source I region has only probed scales much larger than the disk \citep{hir2020,Pattle2021}. For example, Fig. 15 of \citet{hir2020} shows dust polarization vectors mostly on the larger scale of the hot core surrounding Source I.
In this study, we investigate the polarization patterns of the Source I protostellar core at the relatively small scale of the disk and the corresponding outflow-driving region.
First, we execute a high-resolution core collapse simulation using a nested grid code.
Then, we solve the radiative transfer problem and make synthetic polarization maps. 
Finally, we compare the polarization map of our synthetic observation with the SiO line polarization observation of Source I \citep{hir2020}, in order to understand the configuration of the magnetic field around the outflow-driving region, as well as shed light on the polarization mechanism. We examine various tilt angles between the magnetic axis and the plane of the sky.
The synthetic polarization map corresponds to the polarization due to emission from magnetically-aligned elongated dust grains. Hence a direction that is perpendicular to the polarization can be inferred as the direction of the local plane-of-sky magnetic field when averaged along the line of sight. 
In our analysis, the averaging process is weighted by the density of dust, which is proportional to the gas density, and is affected by varying magnetic field directions along the line of sight. Hence, the direction of the polarization segments may not be intuitively predictable. 

We note that \cite{hir2020} measured the magnetic orientation in a different way, through polarization of line emission from SiO. The SiO transitions were interpreted as a mixture of thermal and maser emission. In masers, energy levels are inversely populated, i.e., $n_lg_u-n_ug_l < 0$ where $g_l$ and $n_l$, and $g_u$ and $n_u$ are the statistical weights and population densities of the lower and upper levels, respectively. Therefore, the extinction coefficient ($\kappa^L\propto [n_lg_u-n_ug_l]$) and the optical depth of the line ($\tau^L = \int \kappa^L ds$, where $ds$ is the path of integration) are negative. This leads to the very high peak brightness temperatures that are observed, since the radiative intensity rises exponentially with the optical depth.


Spectral line polarization from masers was first studied theoretically by \cite{GKKa} and \cite{GKKb}. The basic principles are similar to the classical Goldreich-Kylafis (GK) effect \citep{gk81,gk82}.
Here, we describe some of the basic principles of the GK effect. In the presence of a magnetic field, a molecular rotational level splits into magnetic sublevels. The GK effect
predicts a linear polarization of radio-frequency lines, with the polarization being aligned either parallel or perpendicular to the the plane-of-sky (POS) component of the magnetic field. Physically, the GK effect arises when magnetic sublevels are unequally populated. In the opposite case, where magnetic sublevels are equally populated, emission lines would be unpolarized. For the GK effect to arise and for the polarization degree to be sufficient enough to be measured observationally, certain requirements need to be fulfilled: \begin{enumerate*}[label=(\roman*)]
\item the optical depth of the line $\uptau^L$ needs to be moderate ($\sim1$) and anisotropic; \item the radiative transition rates need to be comparable to the collisional rates; and \item the Zeeman splitting must exceed both the collisional frequency and the radiative transition rates.
\end{enumerate*}

The first two of the above requirements need to be satisfied in order for the magnetic sublevels to be unequally populated. Collisional excitation and de-excitation would tend to populate magnetic sublevels at equal rates. The same is true if the radiation is isotropic, due to large optical depth or if the collisional coefficient is dominant in comparison to the radiative transition rates. Finally, the third requirement needs to be satisfied in order for the polarization direction to be either parallel or perpendicular to the magnetic field. The polarization direction is determined from the sign of $n_u - n_\pm$, where $n_\pm$ are the populations of the magnetic sublevels of the upper level. If this condition is not satisfied, the polarization direction would be determined by the orientation of the line-of-sight (LOS) direction with respect to the anisotropic optical depth. 

For masers, the condition that $\tau^L\sim1$ is not necessary since the optical depth can be anisotropic even if $\tau^L\gg1$. Additionally, the high fractional linear polarization observed in many SiO masers can be explained by anisotropic pumping from a central source \citep{Watson2008}.

Regardless of the exact details involved in the unequal population of the magnetic sublevels, our comparison of a synthetic 
polarization map of dust-continuum emission with an observed map of SiO line polarization in the same location can help resolve an ambiguity regarding the magnetic-field direction inferred from the GK effect: whether the induced line polarization is parallel or perpendicular to the local magnetic field direction \citep[see discussion in][]{gk82}. 

We review key features of
both the MHD and radiative transfer simulations in Section~\ref{sec:methods}. We present our dynamical and radiative transfer simulations, and compare our
results with the ALMA SiO polarization measurements from Source I in Section~\ref{sec:results}. Section~\ref{sec:discussion} gives further discussion on the alignment mechanism and whether or not the observation is consistent with the standard paradigm of magnetized collapse. A summary is provided in Section~\ref{sec:summary}.

\section{Methods} \label{sec:methods}
\subsection{Magnetohydrodynamic Simulations}
The collapse of a dense molecular cloud core is followed from the prestellar phase and is evolved into the protostellar phase with nonideal MHD simulations. The initial conditions and numerical techniques and settings are almost identical to that in several previous simulations \citep[e.g.,][]{Machida2012,Matsushita2017,Machida2013,Tomida2017,Machida2020}. Here we highlight some key features. 
A Bonnor-Ebert (BE) density profile is used to construct the initial state of the prestellar core. Fundamental units of time ($1/\sqrt{G\rho_c}$), length ($c_s/\sqrt{G\rho_c}$), and mass ($c_s^3/\sqrt{G^3\rho_c}$) can be constructed from the initial central mass density $\rho_c$, the isothermal sound speed $c_s$, and the gravitational constant $G$. Such models have been previously applied to the case of low-mass star formation \citep[e.g.,][]{machida2008high}, but we can also adapt parameters in order to model initial conditions that are characteristic of high-mass star formation. Here we start with a BE density profile with radius $R_{\rm cl}$ that is twice the critical BE radius. For a choice of central number density $n_c = 9.6 \times 10^4\cc$ (note $n_c \equiv \rho_c/m$, where $m=2.3\, m_H$ is the mean particle mass and $m_H$ is the hydrogen mass) and an isothermal temperature $T=40$\,K ($c_s \equiv (kT/m)^{1/2} = 0.38$ km s$^{-1}$), the total mass is $M_{\rm cl}=40\, M_\odot$ and the outer radius is $R_{\rm cl} = 6.0 \times 10^4$ au. 
A uniform lower density medium is placed outside the initial cloud $r>R_{\rm cl}$ in order to mimic an interstellar medium.

The core is characterized by a dimensionless mass-to-flux ratio (normalized to the critical value $(2\pi G^{1/2})^{-1}$) $\mu_0=2$, a ratio of thermal to gravitational energy $\alpha_0=0.42$ and ratio of rotational to gravitational energy $\beta=0.024$.
The initial cloud density is enhanced by a factor $f$ ($=1.68$) to promote the contraction \citep[see][]{Machida2020}. The initial magnetic field and rotation rate have uniform values consistent with the dimensionless parameters above.
The cloud is partially
ionized and in the densest regions the magnetic field is not well coupled to the neutral material and the flux-frozen approximation is not valid. We solve the resistive MHD equations including self-gravity:
\begin{flalign}
&\frac{\partial \rho}{\partial t} + \nabla \cdot (\rho\, \vec{v}) = 0, \label{eq1}\\
&\rho \frac{\partial\, \vec{v}}{\partial t} + \rho(\vec{v} \cdot \nabla)\,\vec{v} = -\nabla P + \frac{\vec{j}}{c} \times \vec{B} - \rho  \nabla \Phi, \\
&\frac{\partial\, \vec{B}}{\partial t} = \nabla \times (\vec{v} \times \vec{B}) + \eta \nabla^2 \vec{B}, \label{eq3}\\
&\nabla^2 \Phi = 4\pi G \rho, \label{eq4}
\end{flalign}
where $\rho, \vec{v}, P, \vec{B}, \eta, $ and $\Phi$ signify the density, velocity, pressure, magnetic field, resistivity, and gravitational potential, respectively. The electric current density $\vec{j}=c\,(\nabla \times \vec{B})/4\pi$. 
We use the resistivity $\eta$ formulated in \cite{Machida2007,machida2008high} according to \cite{nakano2002mechanism} as 
\begin{equation}
\eta = \dfrac{740}{X_e}\sqrt{\dfrac{T}{10\, {\rm K}}} \left[ 1-{\rm tanh}\left( \dfrac{n}{10^{15}\,{\rm cm}^{-3}}  \right)  \right] \, \, \, {\rm cm}^2\,{\rm s}^{-1},
\label{eq:etadef}
\end{equation}
where $T$ and $n$ are the gas temperature and number density, respectively, and 
\begin{equation}
X_e =  5.7 \times 10^{-4} \left( \dfrac{n}{{\rm cm}^{-3}} \right)^{-1}
\end{equation} 
is the ionization degree of the gas.
In \cite{nakano2002mechanism}, both the ionization degree and resistivity in a collapsing cloud were estimated by solving chemical reactions of charged and neutral species. 
In addition, to solve Equations~(\ref{eq1})--(\ref{eq4}), we use the barotropic equation of state that is modelled from radiation hydrodynamics calculations \citep{Larson1969,Masunaga2000} and is described as
\begin{equation}
P=c_{\rm s}^2 \rho \left[1+\left( \frac{\rho}{\rho_{\rm cri}}\right)^{2/3} \right].
\end{equation}
Here $c_{\rm s}$ is the sound speed of the initial cloud and we set $\rho_{\rm cri}=5\times10^5\rho_{\rm c}$, where $\rho_c$ is the central density of the initial cloud. 
These settings are almost the same as in \citet{Matsushita2017} and \citet{Machida2020}, in which the protostellar outflow in the massive star formation process was investigated.
The MHD equations are solved using a finite-difference method that is second-order accurate in time and space. 

To ensure coverage of the diverse scales associated with the prestellar cloud core ($\sim 10^3 - 10^5$ au) and circumstellar disk ($\sim 1 - 100$ au), these equations are solved on a nested grid \citep{Matsumoto2003,2005MNRAS.362..382M,2005MNRAS.362..369M}.
We prepare 13 different sized grids and represent the grid number by the index $L$ with a range $L=1-13$. 
Each grid has different grid size and cell width, but the same number of cells $(i,j,k) = (64,64,32)$, in which mirror symmetry across the midplane $z=0$ is enforced. 
A finer grid $L=l$ is placed within a coarser grid $L=l-1$ (for a schematic view see Fig. 1 of \citealt{Matsumoto2003}). 
The grid size and cell width are halved with each increment of the grid level.  
The grid size and cell width of the first level ($L=1$) are 
$1.85 \times 10^6$\,au and $2.95 \times 10^4$\,au, respectively, while 
those for the finest level ($L=13$)  are 460 \,au and 7.2\,au,
respectively. 
The initial molecular (or prestellar) cloud core is immersed in the $L=5$ grid, outside which a uniform lower density medium is imposed as stated earlier.
Note that the large area outside the prestellar cloud core ($L=1-4$) is utilized in order to suppress artificial reflection of Alfv\'en waves \citep{Machida2013}. 
The calculation starts with five levels of the grid ($L=1-5$) and a finer grid is automatically generated so as to resolve the Jeans wavelength with at least 16 cells. 
We simulate the core evolution  up to 
$t \sim 240$ kyr, 
corresponding to a central protostellar mass (inside the sink cell) of 
$M \approx 6.5\, M_\odot$.
We impose a sink cell to accelerate the calculation, in which the sink 
threshold density $\approx 10^{11}\cc$ and sink accretion radius $r_{\rm sink}=10$ au are adopted. Our study focuses primarily on two instances
in the evolutionary period, when the protostar mass is $M = 5.7\,M_{\odot}$ and $6.5\,M_{\odot}$, respectively.
\begin{figure*}
\centering
 \includegraphics[width=1\textwidth]{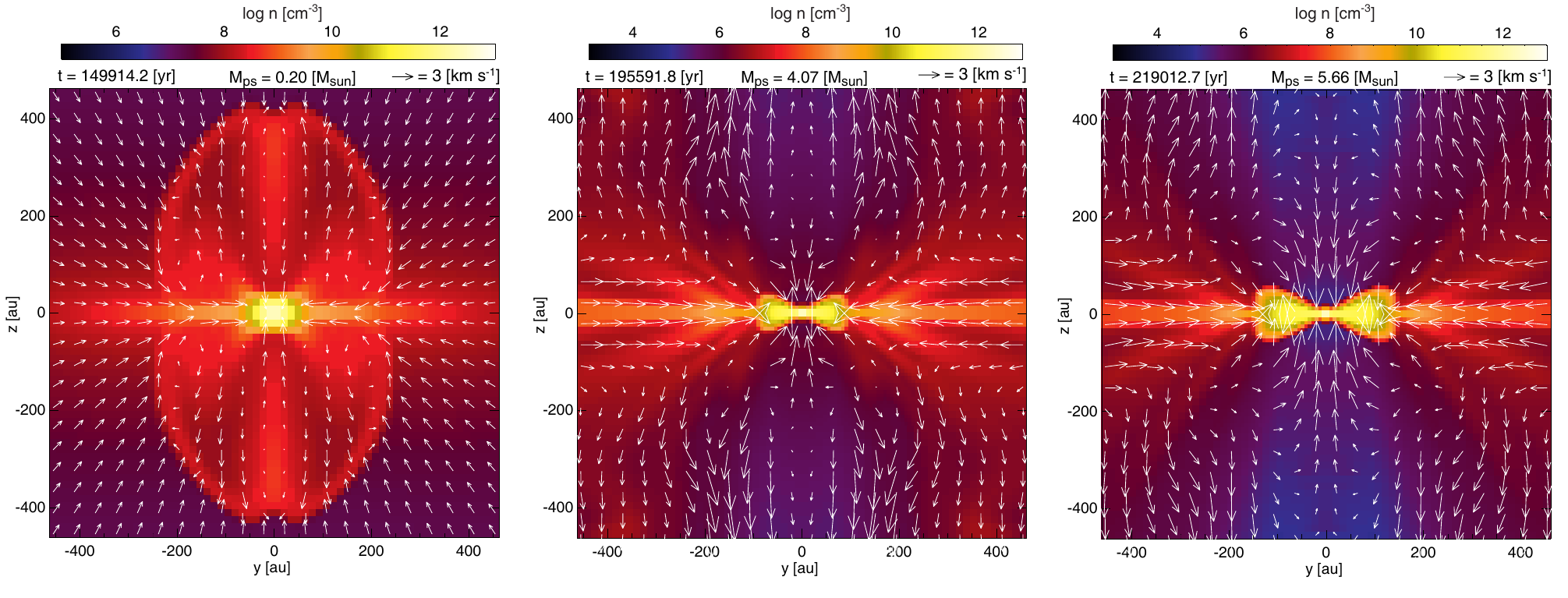}
  \caption{The number density taken along $x = 0$ overlaid with the velocity vectors resulting from the MHD simulation. From left to right the panels are representative of evolutionary stages at which the central mass is $0.2\,M_\odot$, $4.1\,M_\odot$, and $5.7\,M_\odot$, respectively. These represent a fraction of cloud mass in the central protostar of 0.5\%, 10\%, and 14\%, respectively.} \label{mag_dens}
\end{figure*} 

\begin{figure*}
\centering
  \includegraphics[width=1\textwidth]{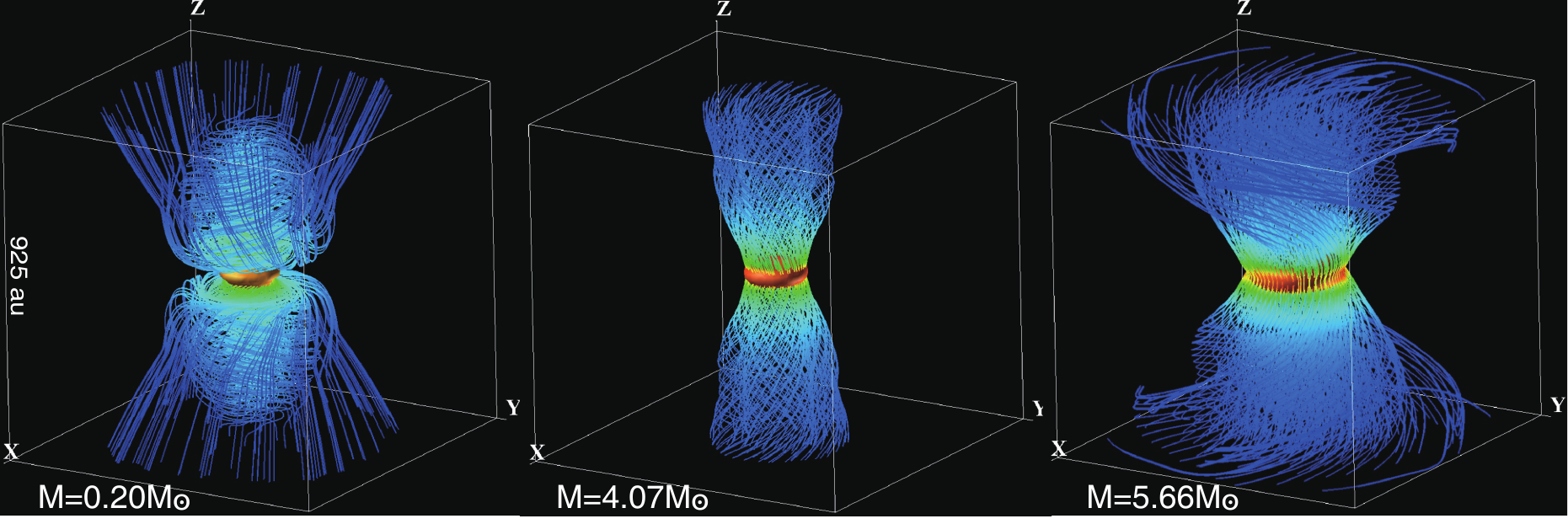}
  \caption{Three-dimensional renderings of the magnetic field lines in our MHD simulation. The lines are shown originating from the surface of the disk, which is an isodensity surface of $n=3 \times 10^{10} \cc$ shown in red. From left to right the panels are representative of evolutionary stages at which the central mass is
  $M = 0.2\,M_{\odot}$,  $4.1\,M_{\odot}$, and $5.7\,M_{\odot}$,
  respectively.} \label{mag_3d}
\end{figure*}
We note that our model does not include radiative feedback from the massive protostar. While this is a limitation in the modeling, our objective in this paper is to fit the highly complex magnetic field line pattern to an observed polarization pattern at scales of $\sim 100-300$ au from the central protostar. In this respect our study is an important step forward.



\subsection{Synthetic Polarization Maps}
In order to simulate synthetic polarization maps, we use the POLARIS code \citep{rei16}, which solves the three-dimensional radiative transfer problem
\begin{align}
    \frac{{\rm d} \vec{S} }{ {\rm d} l } = -\hat{\vec{R}}(\alpha )\,\hat{\vec{K}} \,\hat{\vec{R}}(\alpha)^{-1}\vec{S} + \vec{J}
\end{align}
 for all Stokes parameters contained in the Stokes four-vector $\vec{S} = \left[ I, Q, U, V \right]^{\top}$ (for $I$, $Q$, $U$ and $V$ representing the total intensity, the two states of linear polarization, and the circular polarization, respectively). Here, $\hat{\vec{R}}(\alpha )$ is the rotation matrix, $\hat{\vec{K}}$ is the Muller matrix that describes extinction, absorption and scattering, and $\vec{J}$ is the energy transfer contribution due to emission. We refer the reader to the POLARIS \href{https://www1.astrophysik.uni-kiel.de/~polaris/content/manual.pdf}{software manual} for further details. We model the density, temperature, velocity and magnetic field profiles directly from the MHD simulations. Additionally, we assume that the dust grains are nonspherical (oblate) with a grain size following a power-law distribution \citep{mat77}, with a minimum size $a_{\rm min} = 0.005\, \micro \rm m $ and a maximum size $a_{\rm max} = 2\, \micro \rm m $. The dust-to-gas mass ratio is 0.01 and the grain alignment is taken to follow the alignment mechanism of \cite{DG1951}. The settings of the POLARIS code implementation are similar to that used by \cite{binobasu2021} to model the polarization segments of the prestellar core FeSt 1--457 \citep{kandori2017}. We adopt as a radiation source a protostar positioned at the center of the simulation grid with radius $R = 0.9\, R_{\odot}$ and temperature $T = 4000$ K.
 Although the central high-mass object in Source I may be hotter, little is known about its actual temperature, and we note that our main goal here is to model the polarization directions rather than the polarization intensity.
\section{Results} \label{sec:results}
\subsection{MHD Output}
In Figure \ref{mag_dens}, we show the density of the core along the $x = 0$ midplane, overlaid with the directional velocity field vectors in three instances in its evolutionary period. Furthermore, in Figure \ref{mag_3d} the three-dimensional renderings of the magnetic field are also given about the disk's surface. In the simulation, the outflow is driven by the rotationally-supported disk. 
We can confirm the whole region of the outflow in the left panel of Figure~\ref{mag_dens}. The total size of the outflow at this epoch is about 
900\,au.
Then, the outflow extends up to 
$\sim 6 \times 10^6$\,au
by the end of the simulation. 
In Figure~\ref{mag_dens}, the central yellow region corresponds to the rotationally-supported disk that is enclosed by the pseudodisk colored in orange and purple. 
The rotationally-supported disk gradually grows and has a radius of 
$\sim150$\,au
when the protostellar mass reaches $6.5 M_{\odot}$ (right panel of Fig.~\ref{mag_dens}). The middle and right panel of Figure~\ref{mag_dens} indicate that the outflow is mainly driven by the region near the outer edge of the rotationally-supported disk, as is also seen in a recent ALMA observation \citep{alves17}. The left panel of Figure~\ref{mag_3d} indicates that the configuration of the magnetic field lines is composed of two components. The magnetic field lines are strongly twisted within the outflow, while they have a poloidal configuration outside the outflow (compare left panels of Fig.~\ref{mag_dens} and Fig.~\ref{mag_3d}). The middle and right panels of Figure~\ref{mag_3d} show how the configuration of the magnetic-field lines near the protostar and disk changes as the disk grows in size.
For the magnetic field lines within the outflow plotted in Figure \ref{mag_3d}, the degree of torsion in the middle and right panels is less than that in the left panel. The middle and right panels of Figure \ref{mag_3d} illustrate only the magnetic field lines around the outflow launching region, where the toroidal component is roughly comparable to the poloidal component \citep[e.g.,][]{Konigl00}. In addition, the configuration of the magnetic field lines in the right panel of Figure \ref{mag_3d} is similar to that in the middle panel, while the global torsion, which is produced by the rotation of the large-sized disk, can be seen near the upper and lower boundary of the grid in the right panel. 
As a result, the configuration of the magnetic field lines within the outflow at a particular distance scale will change gradually with time.

\subsection{POLARIS Output}
Our radiative transfer simulations are run for three instances in the core's evolution when the central object has mass $4.1\,M_{\odot}$, $5.7\,M_{\odot}$, and $6.5\,M_{\odot}$, respectively. The simulations are run at a series of declination angles tilted in the direction of the observer. Given independent observational evidence that the disk is seen nearly edge-on \citep{hir2020}, we restrict our parameter space to declination angles $\theta$ in the range of $0^\circ$ to $25^\circ$. By comparing the length scales in the observation to that of our MHD simulations, we find that level $L=12$ is best suited to model the data. POLARIS outputs the polarization segments that result from the integrated scattering and emission properties of the dust grains, where the local emission is perpendicular to the local magnetic field. In order to better gauge the direction of the magnetic field, we have rotated the vectors by $90^\circ$ so that they represent the inferred integrated magnetic field direction.

\subsection{ALMA Data}
\cite{hir2020} 
estimated the magnetic field strength in the outflow lobe of Source I to be approximately 30 mG using the Davis-Chandrasekhar-Fermi method and suggested that the configuration is either toroidal or poloidal depending on how the polarization vectors are oriented relative to the magnetic field. Here we further probe the magnetic field structure and provide insight as to how the polarization vectors may be oriented with respect to the local magnetic field. The polarization segments of the POLARIS output are compared with the observed polarization directions measured from the $J=2-1$ line of SiO \citep[see Fig. 5 and table 1 of][]{hir2020}. In order to assess the alignment quantitatively, we compute the mean residual angle that the polarization vector makes with the $z$-axis. That is, in both the POLARIS simulations and the ALMA observation, the quantity $0\leq \phi \leq \pi /2$ is defined as the offset angle made between the polarization vector and the $z$-axis. Furthermore, the quantity $\Delta \hat{\phi}$ is the absolute residual between the mean of the offset angles from the ALMA observation and the POLARIS simulation. That is,
\begin{equation}
\Delta \hat{\phi}^{(L)} = \left| \hat{\phi}_A - \hat{\phi}_P^{ (L)} \right|\, , \label{abs_res}
\end{equation}
where $\hat{\phi}$ is the mean offset angle given by
\begin{equation}
    \hat{\phi} = \frac{1}{N_P}\sum_i^{N_P} \phi_i \, ,
\end{equation}
in which $N_P$ is the total number of polarization vectors. In equation \eqref{abs_res}, the superscript $(L)$ is taken to refer to the nested grid level in the MHD/POLARIS simulation and the subscripts $A$ and $P$ represent the mean offset angle taken for the ALMA observation and the POLARIS output, respectively. By default, POLARIS will output polarization based on grain alignment perpendicular to the local magnetic field, and so in order to assess the alignment mechanism in the ALMA observation, separate calculations are performed assuming both perpendicular and parallel alignment. The output with the smallest value for $\Delta \hat{\phi}$ can be selected as the most plausible model. 

\begin{figure*}
\centering
  \includegraphics[width=\textwidth]{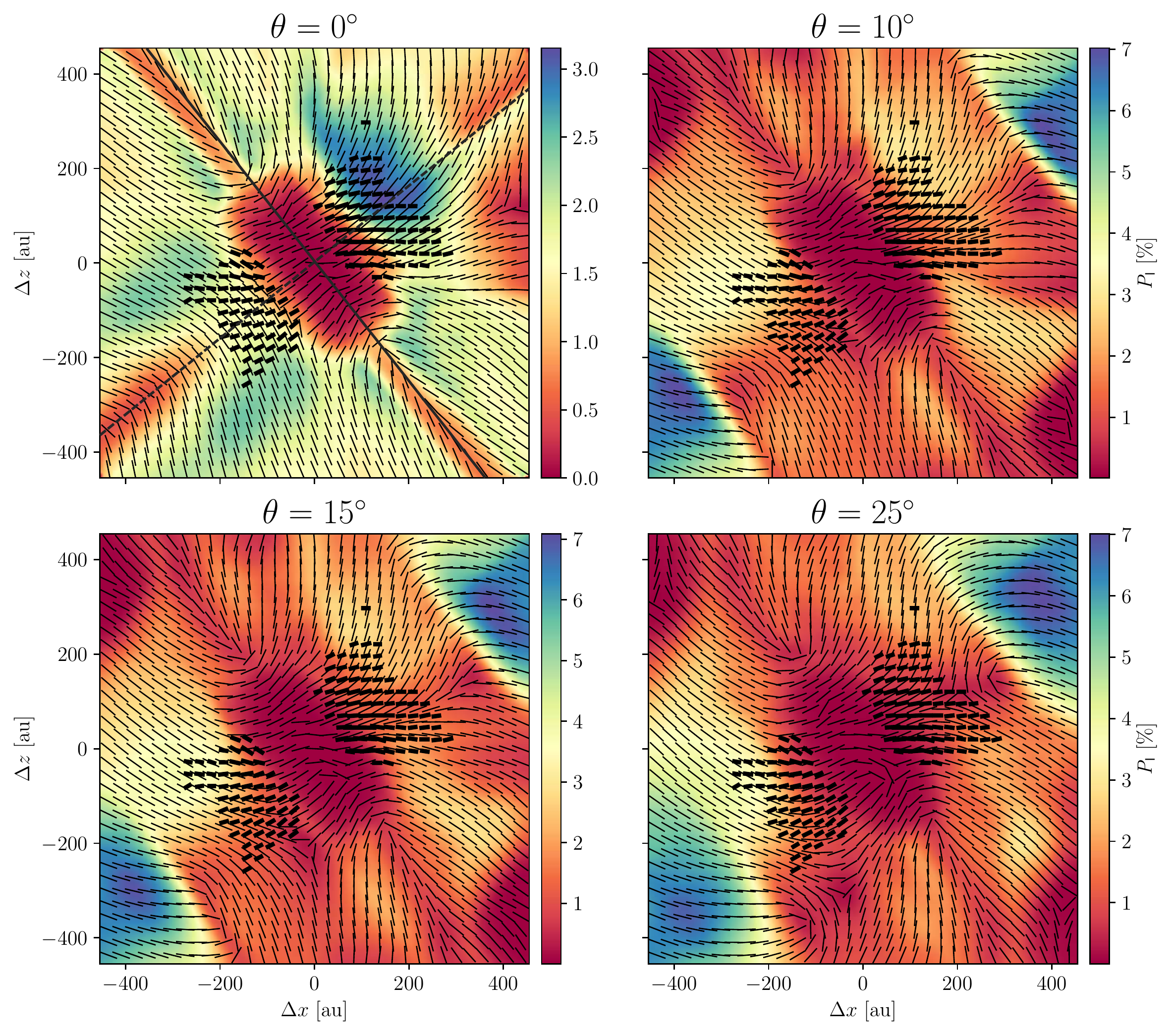}
  \caption{The polarization segments of the ALMA data (bold vectors) overlaid with the simulated synthetic polarization maps from POLARIS. The simulated polarization segments (thin black lines) have been rotated by 90$^{\circ}$ so as to be aligned parallel to the local magnetic field. The colors denote the fractional polarization, with values given in the color bar. The figures are given for the time instance when $M = 6.5\,M_{\odot}$ at $L = 12$. The declination angle $\theta$ is tilted toward the observer. In the top left panel ($\theta = 0^{\circ}$), we include a solid and dashed line to represent the disk midplane and outflow axis, respectively. } \label{alma_pol_1}
\end{figure*}

We illustrate the output polarization vectors from the simulation and their comparison with the observed SiO polarization in Figure~\ref{alma_pol_1}. Here, we show the model at an evolutionary stage with central object mass $M=6.5\msun$ for four different declination angles. 
Note that we rotate the outflow axis of the system by $51^{\circ}$ in the plane of the sky such that it matches the orientation in RA and declination coordinates presented by \cite{hir2020}. However, we also run the horizontal coordinate from left to right, in a mirror image of that presented in \cite{hir2020}, since they had the negative offsets to the right and positive offsets to the left.
In the top left panel of Figure~\ref{alma_pol_1} ($\theta = 0^{\circ}$) we include a solid and a dashed line to represent the direction of the disk midplane and the outflow axis, respectively. The bold black vectors in Figure~\ref{alma_pol_1} are the measured SiO polarization segments of \citet{hir2020}.
The colors represent the fractional polarization estimated by POLARIS. However we do not expect observations to reflect these values as our simulation assumes perfect alignment efficiency and there are many unknowns about the dust emissivity.

\begin{figure}
    \hspace{-0.5cm}
\centering
    \includegraphics[width=0.48\textwidth]{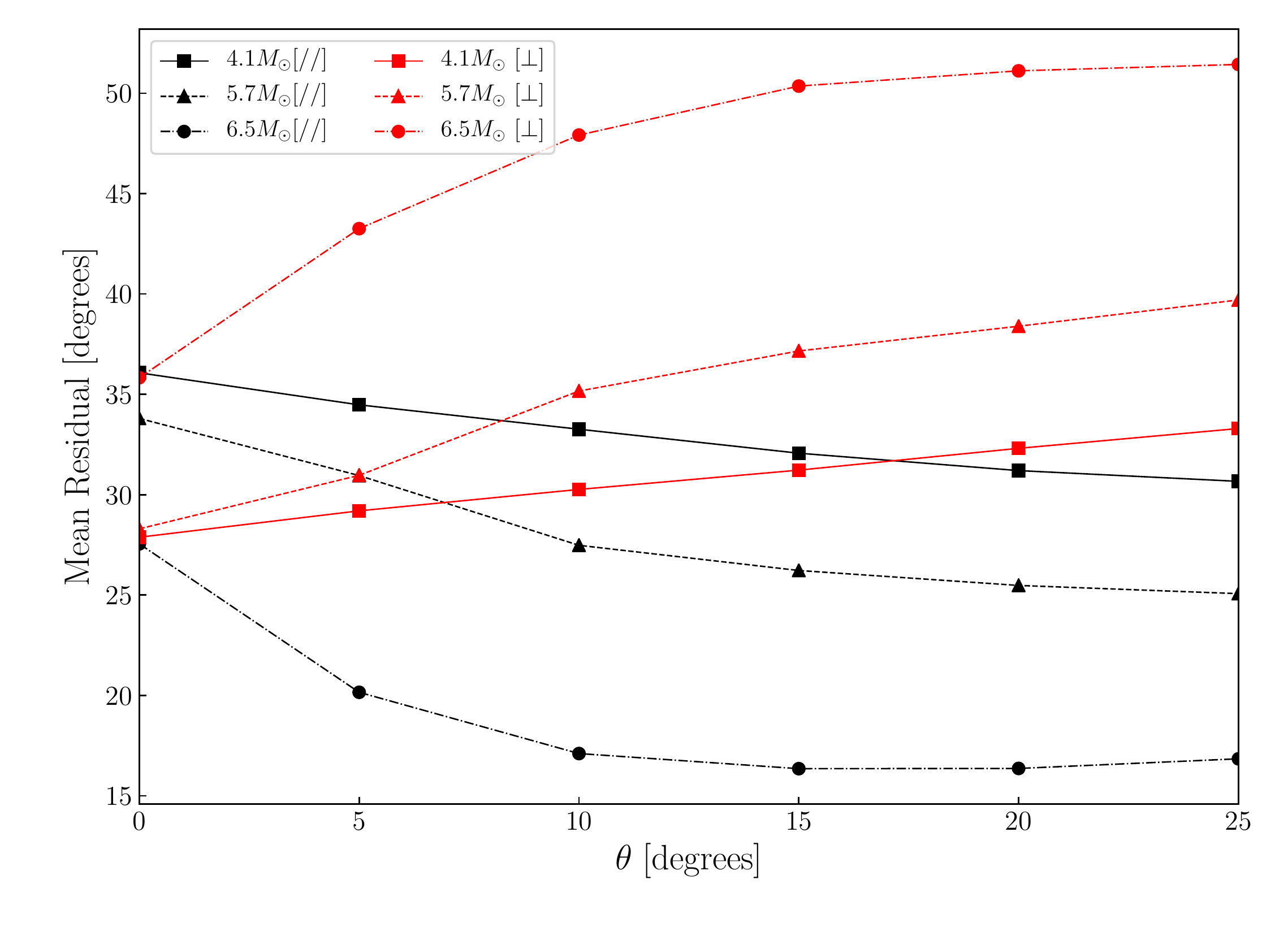}
    \vspace{-0.5cm}
    \caption{Model performance given in terms of the mean residual made between the synthetic polarization segments from the simulation with the observed polarization vectors from ALMA. The metrics are assessed at three instances under different alignment assumptions. Labels tagged $[$ // $]$ assume parallel alignment between magnetic field and polarization vectors, whereas labels tagged $[ \bot ]$ assume perpendicular alignment.}
    \label{fig:moa}
\end{figure}

The results of our calculation on the mean residual angle for the different evolutionary stages, each viewed at multiple declination angles, is shown in Figure~\ref{fig:moa}. We fit the observed polarization vectors considering both possibilities of them being parallel (tagged $[$ // $]$) or perpendicular (tagged $[ \bot ]$) to the local magnetic field. Our results show that the most plausible model (minimum residual $\Delta \hat{\phi}$) is the one at evolutionary stage of $M=6.5\msun$ viewed at $\theta = 15^{\circ}$ and in which the observed vectors are parallel to the local magnetic field.

\begin{figure*}
\centering
  \includegraphics[width=\textwidth]{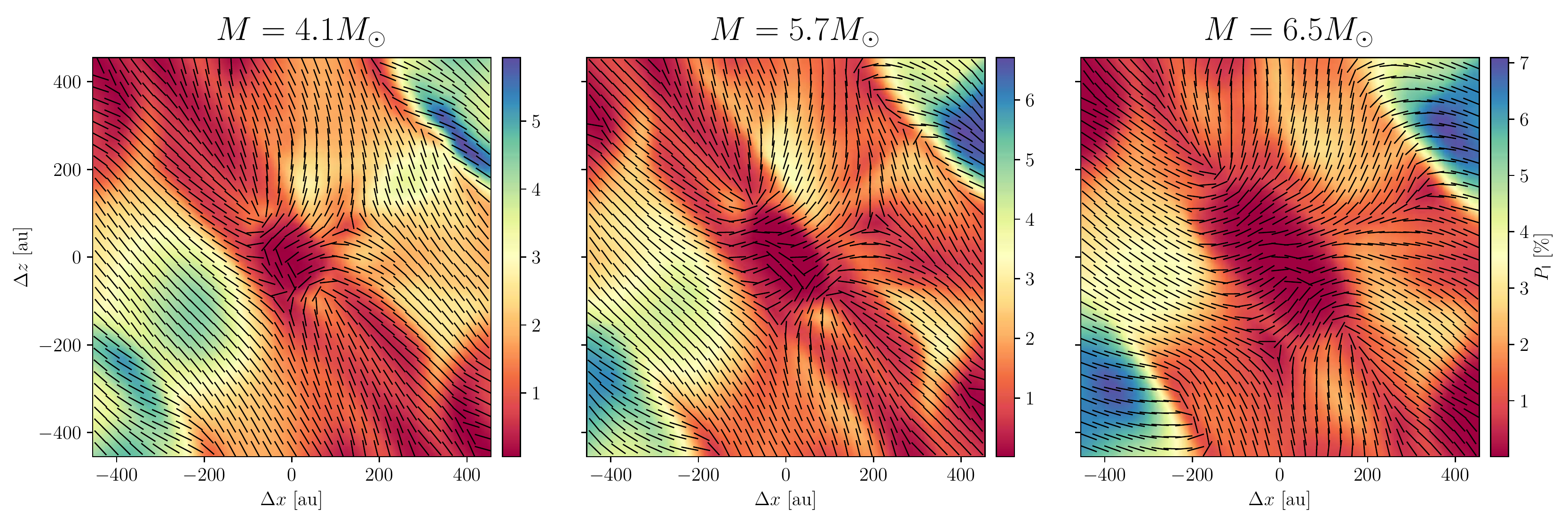}
  \caption{Evolutionary time lapse of the most plausible model given in terms of the central object mass. Each of the figures demonstrate the model at $L=12$ and at a nearly edge-on view with a declination angle of $\theta = 15^{\circ}$. The black vectors and color bar have the same meaning as in Fig.~\ref{alma_pol_1}.  } \label{alma_pol_3}
\end{figure*}

In Figure \ref{alma_pol_3} we show snapshots of the synthetic polarization vectors viewed at $\theta = 15^{\circ}$ at three instances in time. We represent these temporal snapshots in terms of the central object mass. We note that the size of the disk increases as time increases, and the fraction of polarization is weaker along the disk midplane and stronger along the outflow axis.

\begin{figure*}
\centering
  \includegraphics[width=0.9\textwidth]{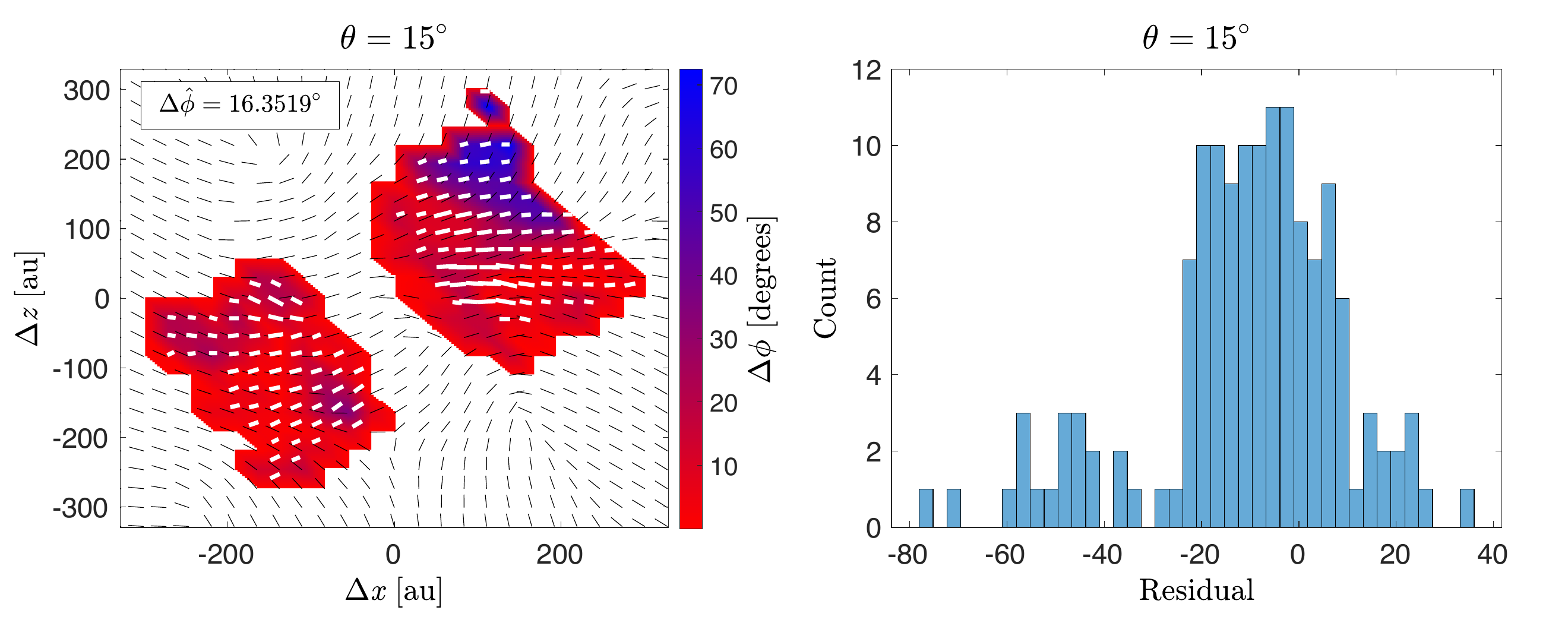}
  \caption{Absolute residual angles made between the ALMA data with the polarization vectors from the POLARIS simulation. The calculations are performed assuming polarization vectors are aligned parallel to the local magnetic field. }
  \label{perf_figs}
\end{figure*}

The performance assessments are given graphically in Figure \ref{perf_figs} where the left panel represents a positional heat map for the residuals made between the simulation and ALMA observations and the left panel summarizes these residuals in a histogram. We note that in these comparisons, no actual fit is performed. That is, there is no form of regression or optimization between the simulation and the observations, but rather just a mere comparison and overlay. The MHD and POLARIS simulations are run completely independent from the ALMA data. As such, we are not concerned with the lack of normality in the distribution of residuals. These performance metrics are used for the sole purpose of assessing the accuracy and representation of our simulations in making inference on the observational data.

\begin{figure*}
\centering
    \includegraphics[width=0.88\textwidth]{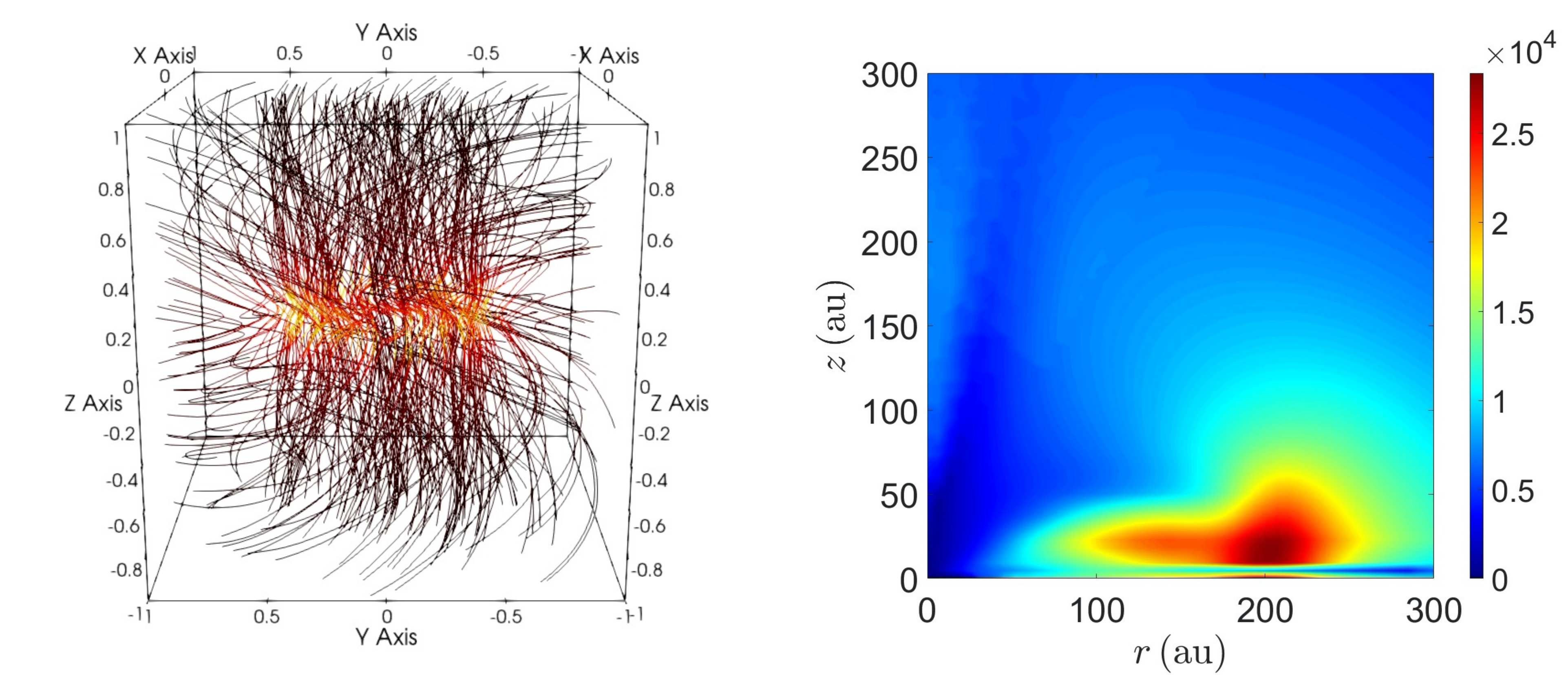}
    \caption{Left: A three-dimensional rendering of the magnetic field lines for $M = 6.5\, M_{\odot}$ and $L = 12$ tilted at $\theta = 15^{\circ}$ toward the observer. Areas of increasing magnetic field strength go from black to red. Right: Azimuthally-averaged magnetic field strength (in $\upmu$G) for $M=6.5\, \msun$ at $L=12$.}
    \label{fig:Bfield_heatmap_3d}
\end{figure*}

The left panel of Figure \ref{fig:Bfield_heatmap_3d} is a three-dimensional rendering of the magnetic field lines at the evolutionary stage $M = 6.5\, M_{\odot}$ at $L = 12$ and tilted $15^\circ$ towards the observer. The rendering illustrates the highly twisted nature of the magnetic field within the outflow zone. The strength of the field is interpreted by a black-red-yellow color gradient, with black being the weakest. 

The right panel of Figure \ref{fig:Bfield_heatmap_3d} shows the strength of the azimuthally-averaged total magnetic field strength in the ($r,z$) plane. The total field strength and the corresponding number density in the outflow region are $B\sim 10 \ {\rm mG}$ and $n_{\rm H_2} \sim 10^4-10^5\, {\rm cm}^{-3}$ (see right panel of Fig.\ref{mag_dens}), respectively. Thus, even without anisotropic radiative pumping from a central source, 
all the conditions required for the GK effect to arise are fulfilled. Specifically, given the values of the collisional and Einstein A coefficients of the ${\rm SiO}$ ($J = 2 - 1$) rotational transition\footnote{https://home.strw.leidenuniv.nl/~moldata/SiO.html.} \citep{leiden}, the collisional and radiative transition rates are comparable. Additionally, assuming that the dipole moment of ${\rm SiO}$ is comparable to the nuclear magneton, the Zeeman splitting is $\sim 2-3$ orders of magnitude greater than the collisional and radiative transition rates.

\section{Discussion} \label{sec:discussion}

The process of high-mass star formation remains elusive because massive YSOs are relatively distant, deeply embedded, and rare because of their rapid evolution \citep[see e.g.,][for a recent review]{Motte2018}.
Despite their rapid formation timescales and strong radiation feedback, a growing amount of observations suggest that high-mass star formation is analogous to a scaled-up version of low-mass star formation.
Here we have employed a scaled-up version of a low-mass MHD model of core collapse. 
Our model is post-processed to fit the Source I data, and we have not picked model parameters specifically to fit the object.
We make a comparison of observed polarization vectors with synthetic polarization segments from multiple time snapshots of the model.

\subsection{Mass of the Central Object}
The highly embedded Source I is actually the nearest example of ongoing high-mass star formation. The central object is obscured and is surrounded by gas in a flattened morphology of size $\sim 100$ au seen in mm continuum emission. SiO maser emissions suggest rotational motion consistent with a nearly edge-on protostellar disk \citep{Reid2007,Ginsburg2018}, and the rotation curves of various molecular lines suggests that the mass of the central object is $\sim 5-8\, M_\odot$ \citep{Kim2008,Plambeck2016}. Using H$_2$O and NaCl emission lines, \cite{Ginsburg2018} inferred an even higher mass of $\sim 15\, M_\odot$. Such a mass is consistent with an idea that Source I and the nearby Becklin–Neugebauer object, along with a third object, Source x, are recoiling from one another after the dynamical decay of a multiple star system. However, the outflow from Source I does not appear to be bowed; the outflow moving with respect to a surrounding medium is expected to be swept back due to ram pressure \citep{Plambeck2009,hirota17}. Here we do not include any external directional ram pressure, so our model is more consistent with an isolated star forming event, as implied by the lack of an observed bow shock structure.

\subsection{Magnetic Field and Polarization Vectors}
Our results demonstrate that an approximately toroidal magnetic field morphology of the outflow zone provides a reasonable fit to the observed SiO polarization pattern in Source I in the outflow zone that is located $\sim 100-300$ au in projected distance from the source. 
Similarly, \citet{lee18} inferred from their observations of the HH 211 protostellar jet that the magnetic field at a distance of $\sim 350 -460 \, {\rm au}$ is mainly toroidal. These studies are consistent with theories of launching mechanisms such as the disk-wind model \citep{Pudritzetal2007diskWind} and X-wind model \citep{Shuetal2000Xwind} that
suggest that the magnetic fields at a distance much greater than the wind/outflow launching radius should be mainly toroidal in order to confine and collimate the flow.

In our modeling, the magnetic axis is aligned with the outflow axis, and we are free to tilt this axis when making the synthetic polarization map. We find that a good fit is obtained when the axis is rotated $15^\circ$ toward the observer. Our model then suggests that the observed SiO line emission polarization vectors are parallel to the local magnetic field. This helps to resolve an ambiguity about the polarization direction.

Previous studies of the GK effect in protostellar outflows via the CO ($J=2-1$) transition \citep{Girart1999, Greaves2001} have also concluded that the spectral line polarization is parallel to the magnetic field. In such situations, the degeneracy between the line polarization being parallel or perpendicular to the magnetic field, can be broken if the angle of the polarization vectors with the polar axis is in the range [35.3$^\circ$, 54.7$^\circ$]$\bigcup$[125.3$^\circ$, 144.7$^\circ$] (see Fig. 5a from \cite{Kylafis1983} for theoretical calculations assuming a one-dimensional velocity field along the polar axis).

\subsection{Radiation Feedback}
We do not include radiative feedback from the protostar in our model, which can be crucial in some phases of massive star formation. We run our model up until the central mass reaches $6.5\msun$, but the main radiative feedback effect is expected to occur at a later stage. \citet[][see also \citet{kui10,kui15}]{kui16} studied the interaction of outflows and radiative feedback using radiation hydrodynamic simulations and found that the effect of radiation pressure on the outflow becomes significant when the central object mass is $M \gtrsim 20-30\,\msun$. At such times the radiation pressure becomes comparable to gravity in the near-circumstellar environment and contributes significantly to the outflow from the central region. In their modeling, \citet{kui16} impose an outflow from the central sink cell at a fixed fraction of the calculated inward mass accretion rate. Future radiation magnetohydrodynamic simulations can better constrain this important issue of the interaction between radiation pressure and the magnetically-driven outflow.

\subsection{Dust-to-gas Ratio}
We have assumed a constant dust-to-gas mass ratio in our simulation. This is generally thought to be a good assumption for the prestellar phase, but in the protostellar phase that contains a protostar, disk, and outflow, it is possible to develop a decoupling of dust and gas in some regions, leading to variable dust-to-gas ratios. \citet{vor18,vor19a} and \citet{vor19b} have shown that significant variations in the dust-to-gas ratio can occur in a protostellar disk. Three-dimensional simulations of the protostellar phase including a magnetic field by \citet[][see also \citet{tsu21}]{leb20} show that significant variations can occur along the direction of the outflow axis, with a dust depletion in the outflow zone.
Such variations of dust-to-gas ratio can affect the ionization state and resistivity and therefore the magnetic field morphology and the emergent polarization properties. The polarization can be affected by the different weightings of different layers along the line of sight if the dust-to-gas ratio varies across layers. These effects of varying dust-to-gas ratio can be explored in a future study.

\section{Summary}
In this study we showed that synthetic polarization maps of the highly twisted magnetic field configuration of an outflow zone in an MHD model can reasonably explain observations of the high-mass star-forming region Orion Source I.
The observations of polarized line emission in SiO show a large scale order in the polarization directions, but they are neither parallel nor perpendicular to the outflow axis. 
Our simulated synthetic maps were used to interpret the SiO line polarization map and led to a better understanding of the morphology and structure of the magnetic field in the outflow region.

Our models show a good fit to the line polarization map of Source I for a protostellar core with central object of mass 6.5 $M_\odot$ and its surrounding disk and outflow that are tilted at an angle 15$^\circ$ towards the observer.
We found that the observed SiO line polarization is most likely parallel to the ambient magnetic field, thereby resolving an ambiguity about its direction.



    
    
    

\label{sec:summary}

\begin{acknowledgments}
We thank the former Department of Applied Mathematics at U.W.O. for providing a home to several of us and enabling the collaborative work that led to this paper. We also thank the referee for comments that helped to improve this manuscript.
A.T. was supported by the Natural Sciences and Engineering Research Council of Canada (NSERC), [funding reference \# CITA 490888-16]. S.B. was supported by a Discovery Grant from NSERC.
\end{acknowledgments}

\bibliography{manuscript}{}
\bibliographystyle{aasjournal}
\end{document}

%% file: top-apj.tex
\newcommand\apjcls{1}
\newcommand\aastexcls{2}
\newcommand\othercls{3}

\newcommand\papercls{\aastexcls}

\if\papercls \apjcls
\usepackage{apjfonts}
\else\if\papercls \othercls
\usepackage{epsfig}
\usepackage{margin}
\usepackage{times}
\fi\fi
\usepackage{ifthen}
\usepackage{natbib}
\usepackage{amssymb, amsmath}
\usepackage{appendix}
\usepackage{etoolbox}
\usepackage[T1]{fontenc}
\usepackage{paralist}

\if\papercls \apjcls
\newcommand\aas{\ref@jnl{AAS Meeting Abstracts}}
\newcommand\dps{\ref@jnl{AAS/DPS Meeting Abstracts}}
\newcommand\maps{\ref@jnl{MAPS}}
\else\if\papercls \othercls
\usepackage{astjnlabbrev-jh}
\fi\fi

\if\papercls \aastexcls
\hypersetup{citecolor=blue, 
            linkcolor=blue, 
            menucolor=blue, 
            urlcolor=blue}  
\else
\usepackage[
bookmarks=true,           
bookmarksnumbered=true,   
colorlinks=true,          
citecolor=blue,           
linkcolor=blue,           
menucolor=blue,           
urlcolor=blue,            
linkbordercolor={0 0 1},  
pdfborder={0 0 1},
frenchlinks=true]{hyperref}
\fi

\if\papercls \othercls

\else

\fi

\providecommand{\adsurl}[1]{\href{#1}{ADS}}

\makeatletter
\patchcmd{\NAT@citex}
  {\@citea\NAT@hyper@{%
     \NAT@nmfmt{\NAT@nm}%
     \hyper@natlinkbreak{\NAT@aysep\NAT@spacechar}{\@citeb\@extra@b@citeb}%
     \NAT@date}}
  {\@citea\NAT@nmfmt{\NAT@nm}%
   \NAT@aysep\NAT@spacechar\NAT@hyper@{\NAT@date}}{}{}

\patchcmd{\NAT@citex}
  {\@citea\NAT@hyper@{%
     \NAT@nmfmt{\NAT@nm}%
     \hyper@natlinkbreak{\NAT@spacechar\NAT@@open\if*#1*\else#1\NAT@spacechar\fi}%
       {\@citeb\@extra@b@citeb}%
     \NAT@date}}
  {\@citea\NAT@nmfmt{\NAT@nm}%
   \NAT@spacechar\NAT@@open\if*#1*\else#1\NAT@spacechar\fi\NAT@hyper@{\NAT@date}}
  {}{}
\makeatother

\makeatletter
\DeclareRobustCommand{\lowcase}[1]{\@lowcase#1\@nil}
\def\@lowcase#1\@nil{\if\relax#1\relax\else\MakeLowercase{#1}\fi}
\makeatother

\DeclareSymbolFont{UPM}{U}{eur}{m}{n}
\DeclareMathSymbol{\umu}{0}{UPM}{"16}
\let\oldumu=\umu
\renewcommand\umu{\ifmmode\oldumu\else\math{\oldumu}\fi}
\newcommand\micro{\umu}
\if\papercls \othercls

\else

\fi

\let\oldsim=\sim
\renewcommand\sim{\ifmmode\oldsim\else\math{\oldsim}\fi}
\let\oldpm=\pm
\renewcommand\pm{\ifmmode\oldpm\else\math{\oldpm}\fi}
\newcommand\by{\ifmmode\times\else\math{\times}\fi}

\newbox{\wdbox}
\renewcommand\c{\setbox\wdbox=\hbox{,}\hspace{\wd\wdbox}}
\renewcommand\i{\setbox\wdbox=\hbox{i}\hspace{\wd\wdbox}}

\newcount\timect
\newcount\hourct
\newcount\minct
\newcommand\now{\timect=\time \divide\timect by 60
         \hourct=\timect \multiply\hourct by 60
         \minct=\time \advance\minct by -\hourct
         \number\timect:\ifnum \minct < 10 0\fi\number\minct}

\catcode`@=11

\newcommand\comment[1]{}

\newcommand\commenton{\catcode`\%=14}

\renewcommand\math[1]{$#1$}
\newcommand\mathshifton{\catcode`\$=3}

\let\atab=&
\newcommand\atabon{\catcode`\&=4}

\let\oldmsp=\sp
\let\oldmsb=\sb
\def\sp#1{\ifmmode
           \oldmsp{#1}%
         \else\strut\raise.85ex\hbox{\scriptsize #1}\fi}
\def\sb#1{\ifmmode
           \oldmsb{#1}%
         \else\strut\raise-.54ex\hbox{\scriptsize #1}\fi}
\newbox\@sp
\newbox\@sb
\def\sbp#1#2{\ifmmode%
           \oldmsb{#1}\oldmsp{#2}%
         \else
           \setbox\@sb=\hbox{\sb{#1}}%
           \setbox\@sp=\hbox{\sp{#2}}%
           \rlap{\copy\@sb}\copy\@sp
           \ifdim \wd\@sb >\wd\@sp
             \hskip -\wd\@sp \hskip \wd\@sb
           \fi
        \fi}
\def\msp#1{\ifmmode
           \oldmsp{#1}
         \else \math{\oldmsp{#1}}\fi}
\def\msb#1{\ifmmode
           \oldmsb{#1}
         \else \math{\oldmsb{#1}}\fi}

\def\supon{\catcode`\^=7}

\def\subon{\catcode`\_=8}

\def\supsubon{\supon \subon}

\newcommand\actcharon{\catcode`\~=13}

\newcommand\paramon{\catcode`\#=6}

\comment{And now to turn us totally on and off...}

\newcommand\reservedcharson{ \commenton  \mathshifton  \atabon  \supsubon 
                             \actcharon  \paramon}

\catcode`@=12
\reservedcharson

\if\papercls \apjcls

\else

\fi

\newcommand\chisq{\ifmmode{\chi\sp{2}}\else\math{\chi\sp{2}}\fi}
\newcommand\redchisq{\ifmmode{ \chi\sp{2}\sb{\rm red}}
                    \else\math{\chi\sp{2}\sb{\rm red}}\fi}
\newcommand\Teq{\ifmmode{T\sb{\rm eq}}\else$T$\sb{eq}\fi}
\newcommand\mjup{\ifmmode{M\sb{\rm Jup}}\else$M$\sb{Jup}\fi}
\newcommand\rjup{\ifmmode{R\sb{\rm Jup}}\else$R$\sb{Jup}\fi}
\newcommand\msun{\ifmmode{M\sb{\odot}}\else$M\sb{\odot}$\fi}
\newcommand\rsun{\ifmmode{R\sb{\odot}}\else$R\sb{\odot}$\fi}
\newcommand\mearth{\ifmmode{M\sb{\oplus}}\else$M\sb{\oplus}$\fi}
\newcommand\rearth{\ifmmode{R\sb{\oplus}}\else$R\sb{\oplus}$\fi}

\renewcommand{\vec}[1]{\boldsymbol{#1}} 